\documentclass[usenatbib]{mn2e}

\usepackage{graphicx}
\usepackage{lscape}

\title{AX\,J1749.1$-$2733 and AX\,J1749.2$-$2725 --- the close pair of X-ray
  pulsars behind the Galactic Center: an optical identification}

\author[Karasev, Lutovinov \& Burenin]
{D.~I.~Karasev$^{1}$\thanks{E-mail:dkarasev@iki.rssi.ru},
  A.~A.~Lutovinov$^{1}$, R.~A.~Burenin$^{1}$\\
  $^{1}$Space Research Institute, Profsoyuznaya str. 84/32, Moscow 117997,
  Russia}

\date{Accepted .... Received ...}
\begin{document}
\pagerange{\pageref{firstpage}--\pageref{lastpage}} \pubyear{2010}

\maketitle

\label{firstpage}

\begin{abstract}
 Two faint X-ray pulsars, AX\,J1749.2$-$2725 and AX\,J1749.1$-$2733, located
 in the direction to the Galactic Center, were studied in detail using data
 of INTEGRAL, XMM-\emph{Newton} and Chandra observatories in X-rays, the
 SOFI/NTT instrument in infrared and the RTT150 telescope in optics. X-ray
 positions of both sources were determined with the uncertainty better than
 $\sim1\arcsec$, that allowed us to identify their infrared counterparts.
 From the subsequent analysis of infrared and optical data we conclude that
 counterparts of both pulsars are likely massive stars of B0-B3 classes 
 located behind the Galactic Center at distances of 12-20 kpc, depending on 
 the type, probably in further parts of galactic spiral arms. In addition, 
 we investigated the extinction law towards the galactic bulge and found 
 that it is significantly different from standard one.

\end{abstract}


\section{Introduction}

Heavily obscured ($N_H\sim10^{23}$ cm$^{-2}$) X-ray source
AX\,J1749.2$-$2725 was discovered by the ASCA observatory in March 1995 and
localized with coordinates R.A.=17$^h$49$^m$10.$^s$1,
Dec.=-27\fdg25\arcmin16\arcsec\ (all coordinates here and below are given in
the J2000 system) and uncertainty about 50\arcsec\ \citep{Tor98}. Authors also
found coherent pulsations with a period of $\sim220.4$~s and
proposed that AX\,J1749.2$-$2725 is an X-ray pulsar in a high-mass X-ray
binary (HMXB) system. These results and conclusions were confirmed later by
\citet{Sak02} during ASCA observations in 1998. No optical
identification of this source was made to date.

The other X-ray source AX\,J1749.1$-$2733, located just a few arcminutes
away from AX\,J1749.2$-$2725, was discovered on Sept 19, 1996 by the ASCA
observatory at R.A.=17$^h$ 49$^m$ 10$^s$,
Dec.=-27\fdg 33\arcmin 14\arcsec\ with the uncertainty of $\sim55$\arcsec\
\citep{Sak02}. Observations of the source with the INTEGRAL observatory
during the flare in September 2003 \citep{greb07} and with the
XMM-\emph{Newton} observatory in March 2007 allowed us to discover a periodicity of X-ray flux from the source with the period of $\sim$132~s \citep{Kar07}\footnote{Note, 
that the pulse period can be half as much due to a double-peaked pulse profile}. The
combination of this result with results of the analysis of ASCA,
XMM-\emph{Newton} and SWIFT spectral data, which revealed a strong
absorption, led us to the conclusion that AX J1749.1$-$2733 is a new X-ray
pulsar in the high-mass X-ray binary (HMXB) system \citep{Kar08}.
Nevertheless the question about the optical counterpart in the system was
still open.

Using XMM-\emph{Newton} data \citet{Zur08} obtained an accurate X-ray
position of AX\,J1749.1$-$2733 R.A.=17$^h$49$^m$06.$^s$8 and
Dec.=-27\fdg32\arcmin32\arcsec.5 with the uncertainty of 2\arcsec. Subsequent
observations of this field of the sky with the ESO/NTT telescope in infrared and
optical bands showed that all three relatively bright stars, suggested by
\citet{Rom07} as possible counterparts of AX\,J1749.1$-$2733, can be ruled
out, but several faint stars remained. One of them (\#5 in the paper of
Zurita Heras and Chaty, 2008) was proposed as a companion star of a possible
Be-nature at a distance of $>8.5$ kpc.

In this paper we reanalyzed XMM-\emph{Newton} data, obtained on March, 31
2007 (Obs.ID 0510010401) during observations of AX J1749.1$-$2733, and found
that the X-ray pulsar AX\,J1749.2$-$2725 was also significantly detected at
the position R.A.=17$^h$49$^m$12$^s$.4, Dec.=-27\fdg25\arcmin37.\arcsec8
with the uncertainty of $\sim2$\arcsec. Combining XMM-\emph{Newton}
measurements with Chandra and SOFI/NTT data we identified IR conterparts for
both X-ray sources.

Observations and data analysis are described in Sect. 2; results of the
X-ray data analysis for AX J1749.2$-$2725 are presented in Sect. 3, followed
by the identification results using SOFI/NTT data from the ESO public
archive and optical data from the RTT150 telescope (Sect. 4 and 5); a brief
summary is given in Sect 6.

\begin{figure}

  \includegraphics[width=\columnwidth,bb=50 380 570
  710,clip]{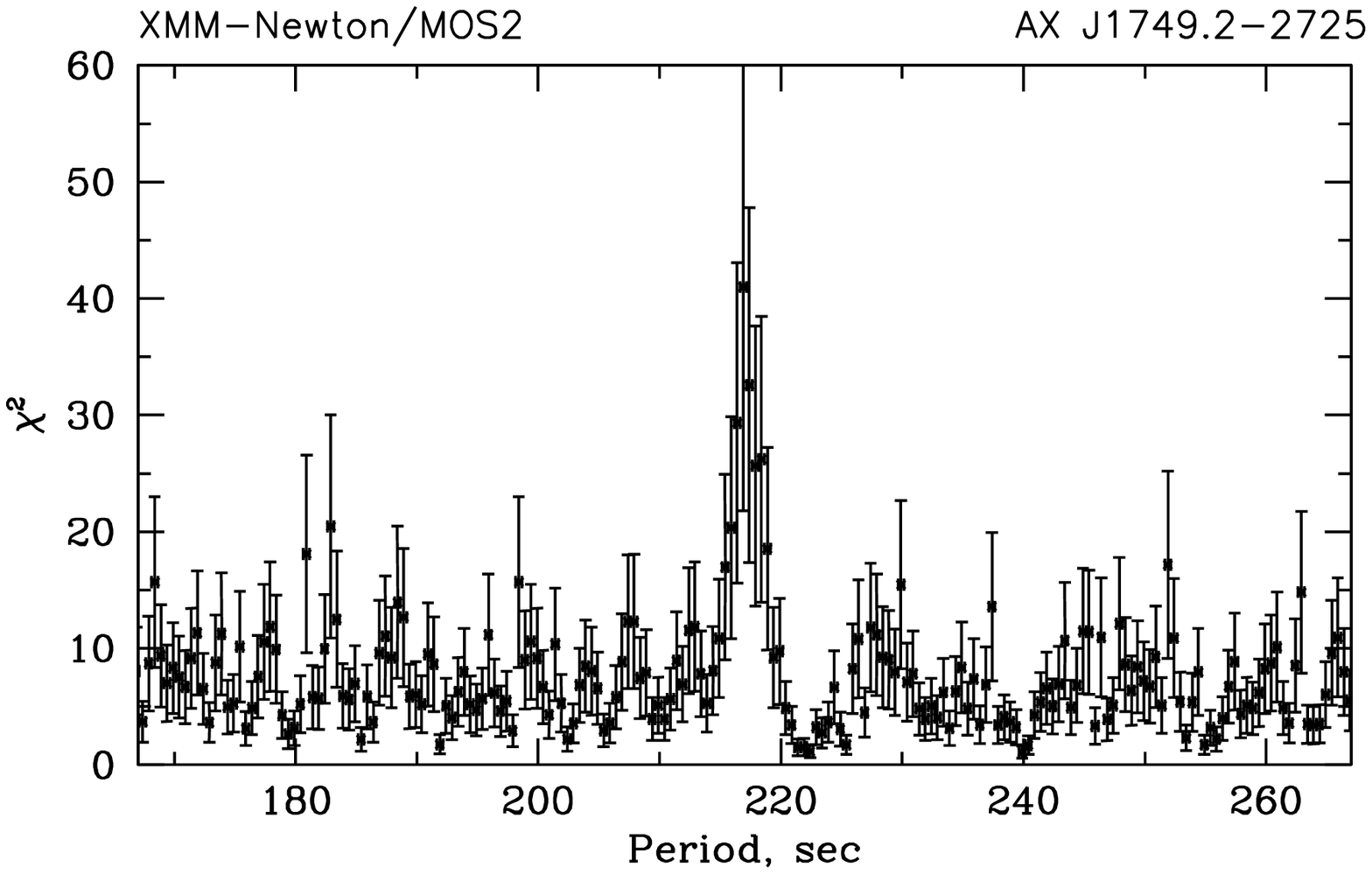}
  \includegraphics[width=\columnwidth,bb=50 380 570
  710,clip]{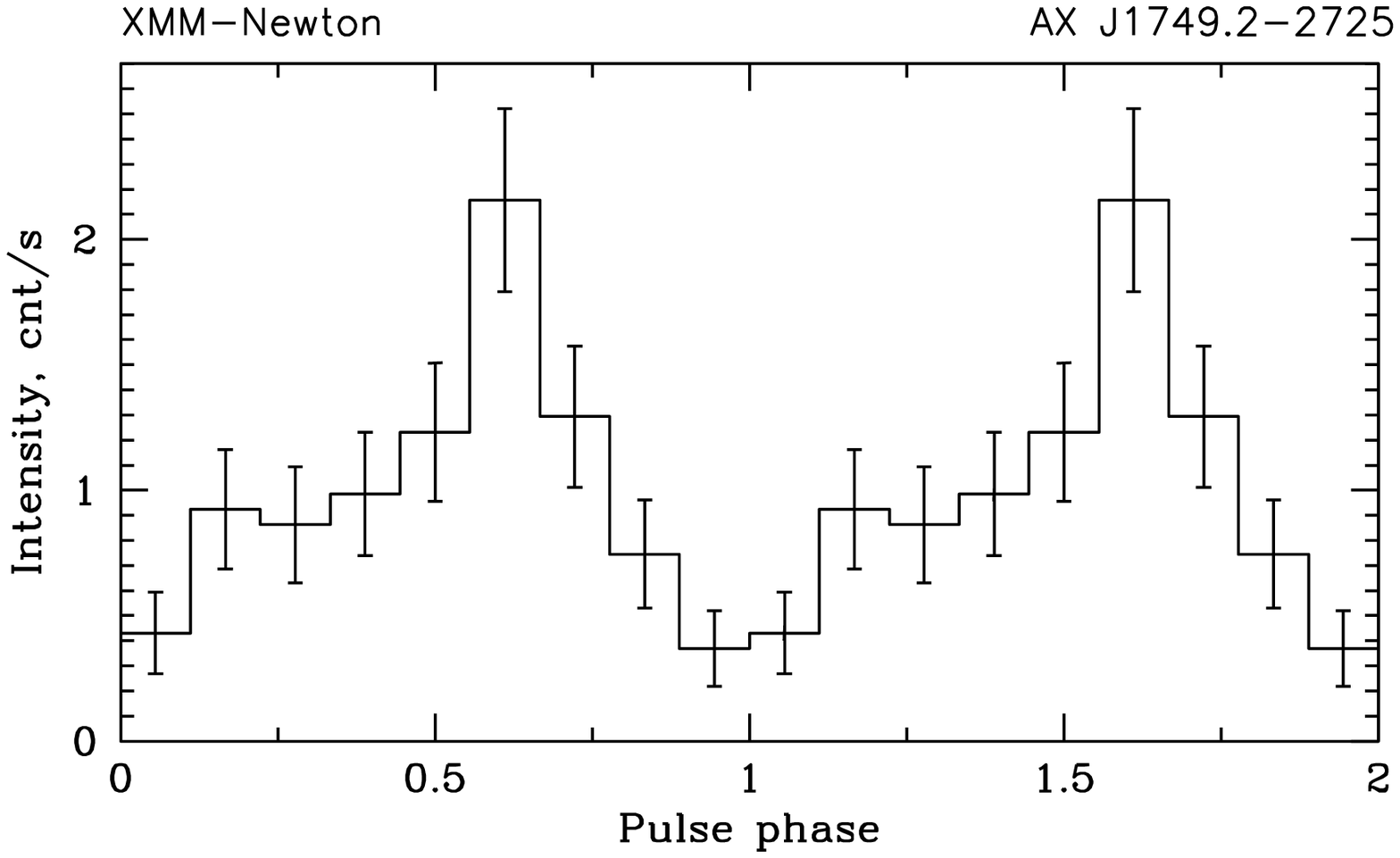}

\caption{AX\,J1749.2$-$2725 periodogram ({\it top panel}) and pulse profile
({\it bottom panel}) reconstructed from the XMM-\emph{Newton}/MOS2 data in
the 2-10 keV energy band.}\label{period}

\end{figure}
\begin{figure}
{
\includegraphics[width=\columnwidth,bb=40 210 570 700,clip]{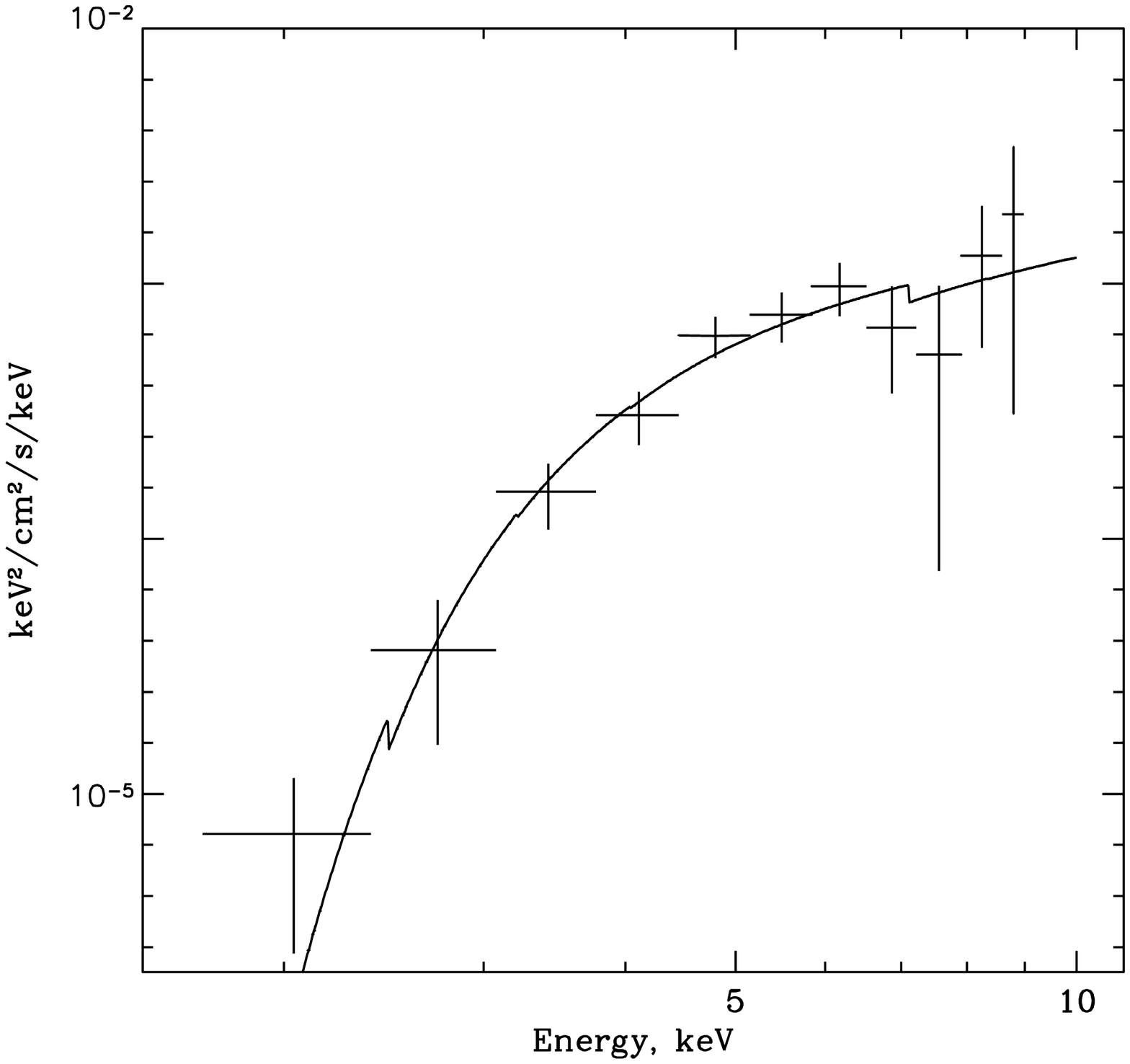}

\vspace {-43mm}
\hspace {42mm}
\includegraphics[width=0.42\columnwidth,bb=40 200 570 700,clip]{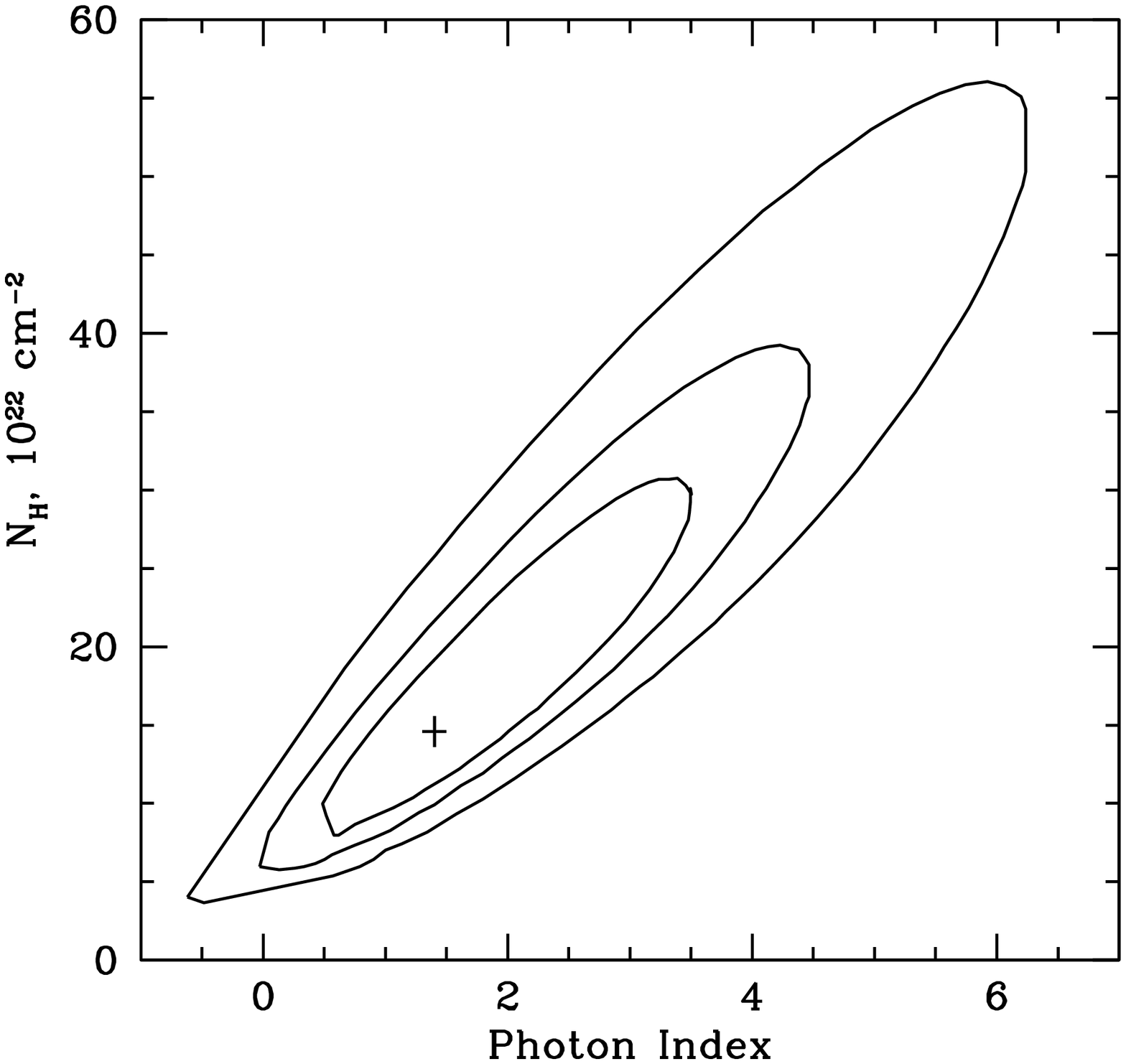}
}
\vspace {10mm}
  \caption{Spectrum of AX\,J1749.2$-$2725 measured with
  XMM-\emph{Newton}/MOS2 in March 2007. Solid line represents a best-fit
  model (main). Confidence contours for proposed model (inset).}  \label{spectrum}

\end{figure}

\section{Observations and Data Analysis}

We analyzed data from the MOS2 camera onboard the XMM-\emph{Newton}
observatory, with a total exposure of $\sim6$~ks, using the standard
XMM-\emph{Newton} software --- Science Analysis System (SAS) v7.0.0. The
final timing and spectral analysis was carried out with the HEASOFT~6.4
package.

For the crosscheck of positional measurements of XMM-\emph{Newton} and final
conclusions about the source coordinates we used results of the pipeline
analysis of the Chandra observatory data, obtained during observations of
AX\,J1749.1$-$2733 and AX\,J1749.2$-$2725 on Febrary 8, 2008 (Obs.ID 9013)
and April 27, 2008 (Obs.ID 7504), respectively.

In order to identify optical counterparts for studied sources we used IR
data of the SOFI instrument of the NTT telescope from the public ESO
archive\footnote{http://archive.eso.org/}. Observations of AX\,J1749.2$-$2725
were performed on March 20, 2001 with the pixel scale of 0.144\arcsec
and total exposures of 4$\times$15 s, 4$\times$15 s and 36+5$\times$15 s in
J, H and Ks bands, respectively; observations of AX\,J1749.1$-$2733 were
performed on April 3-4, 2007 with the pixel scale of 0.288\arcsec and total
exposures of 9$\times$47.3 s and 383$\times$5.9 s in $H$ and $Ks$ bands, respectively.
Additional images in the optical $i\arcmin$-band were obtained with the Russian-Turkish
1.5-m Telescope (RTT150, TUBITAK National Observatory, Antalya, Turkey). The
infrared and optical data were reduced and analyzed using the standard packages
(\emph{IRAF}\footnote{http://iraf.noao.edu/},
\emph{zhtools}\footnote{http://hea-www.harvard.edu/RD/zhtools/},
\emph{WCStools}\footnote{http://tdc-www.harvard.edu/wcstools/}, etc.) and
our own software. The photometry was performed by fitting the point spread
function with the DAOPHOT III
software\footnote{http://www.eso.org/sci/data-processing/software/scisoft/}.  The
photometric solutions for SOFI/NTT images of AX\,J1749.1$-$2733 were taken
from \citet{Zur08} and the solutions for AX\,J1749.2$-$2725 images were
obtained using data of standard stars observations from the ESO archive.
The astrometric solutions of images were found using the \emph{2MASS}
catalog as a reference.

\begin{figure}
  \includegraphics[width=0.9\columnwidth]{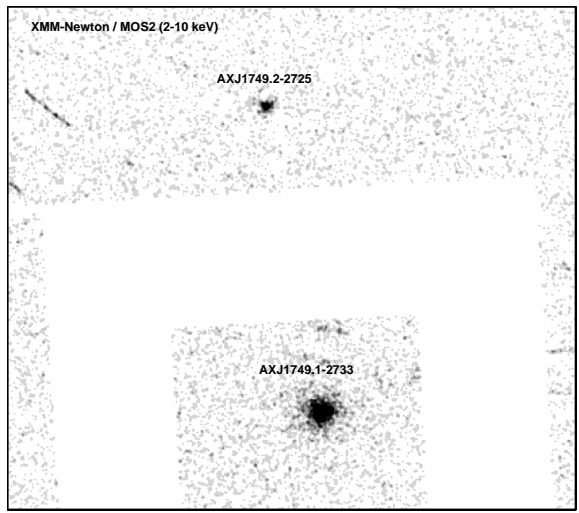}

\caption{The XMM-\emph{Newton}/MOS2 X-ray image (the 2--10~keV energy band)
of the field of the sky near X-ray pulsars AX\,J1749.1$-$2733 and
AX\,J1749.2$-$2725.}\label{maps}

\end{figure}

\section{AX\,J1749.2$-$2725 in X-rays with XMM-\emph{Newton}}

As it was mentioned above, during observations of AX\,J1749.1$-$2733 by
XMM-\emph{Newton} on March 31, 2007 the other X-ray pulsar
AX\,J1749.2$-$2725 was detected in the same field of view. The source was identified
due to X-ray pulsations with a period of $\sim$220~s, which were discovered
earlier by the ASCA observatory \citep{Tor98}. The periodogram of the
AX\,J1749.2$-$2725 background subtracted emission, obtained by
XMM-\emph{Newton}/MOS2 in the 2--10~keV energy band, is shown in
Fig.\ref{period}. Despite of the source faintness a significant excess of the
$\chi^2$-distribution near $\sim$220~s is clearly seen. The best-fit value
of the pulse period is equal to $\sim$216.86$\pm$0.14 s; the uncertainty
corresponds to $1\sigma$ and was determined by the bootstrap method from the
analysis of a large number of simulated light curves \citep[see][for
details]{Tsy05}. The source pulse profile folded with this period is
presented in Fig.\ref{period} (bottom); it has a single--peaked shape and
pulse fraction\footnote{Its value was calculated as $PF = (I_{max}-I_{min}
  )/(I_{max}+I_{min})$, where $I_{min}$ and $I_{max}$ are
  background-corrected count rates at the pulse profile minimum and maximum}
of $\sim$70\% in the 2--10~keV energy range.

The spectrum of the source can be well approximated by a power law model
with the photon index $\Gamma=1.41^{+0.75}_{-1.06}$, modified by the
photoelectric absorption with $N_H=14.1^{+6.13}_{-7.96} \times10^{22}$
cm$^{-2}$ (reduced $\chi^2$ = 0.73 for 18 d.o.f., Fig.\ref{spectrum}). The
corresponding unabsorbed source flux in the $2-10$~keV energy band was
$\sim2.6\times 10^{-12}$ ergs cm$^{-2}$ s$^{-1}$ during XMM-\emph{Newton}
observations, which is about an order of magnitude lower than one measured
by the ASCA observatory in 1995 \citep{Tor98}.  Nevertheless, the spectrum
slope and interstellar absorption are formally in a good agreement
determined from these observations. This fact may by explaned by the large
uncertainty of the determination of source spectral parameters due to the
source faintness. The corresponding confidence contours for the spectral
index and absorption are shown in Fig.\ref{spectrum} (inset).  Nevertheless,
despite of the large uncertainty, the measured value of $N_H$ is
significantly higher than the value of the interstellar absorption in the
direction to the source \citep{Schl98}, which indicates the presence of the
strong intrinsic absorption in the binary system.

The comparison of our pulse period measurements with those of \citet{Tor98}
and \citet{Sak02} points to the acceleration of a neutron star rotation
since the ASCA observations with an average rate of $\dot
P/P\simeq1.3\times10^{-3}$ yr$^{-1}$. Using the expression for the maximum
possible acceleration from \citet{Lip81} we can estimate roughly the source
luminosity as $L_{X,2-100 keV}\simeq2.1\times10^{35}$\,erg~s$^{-1}$ (here we
consider $L_{X,2-100 keV}$ - X-ray luminosity in the $2-100$ keV energy
band, as bolometric). Applying the correction factor of $\sim3$ to make a
transition from the luminosity in $2-100$ keV to the $2-10$ keV energy band
(this factor was derived from the spectral analysis of a large number of
X-ray pulsars in a wide energy band, \citep[see, e.g.][]{Fil05}) we obtain
$L_{X,2-10 keV}\simeq7\times10^{34}$ ergs s$^{-1}$ and the distance to the
source $d\simeq16$ kpc.

\begin{figure*}
  {
    \includegraphics[width=0.9\columnwidth]{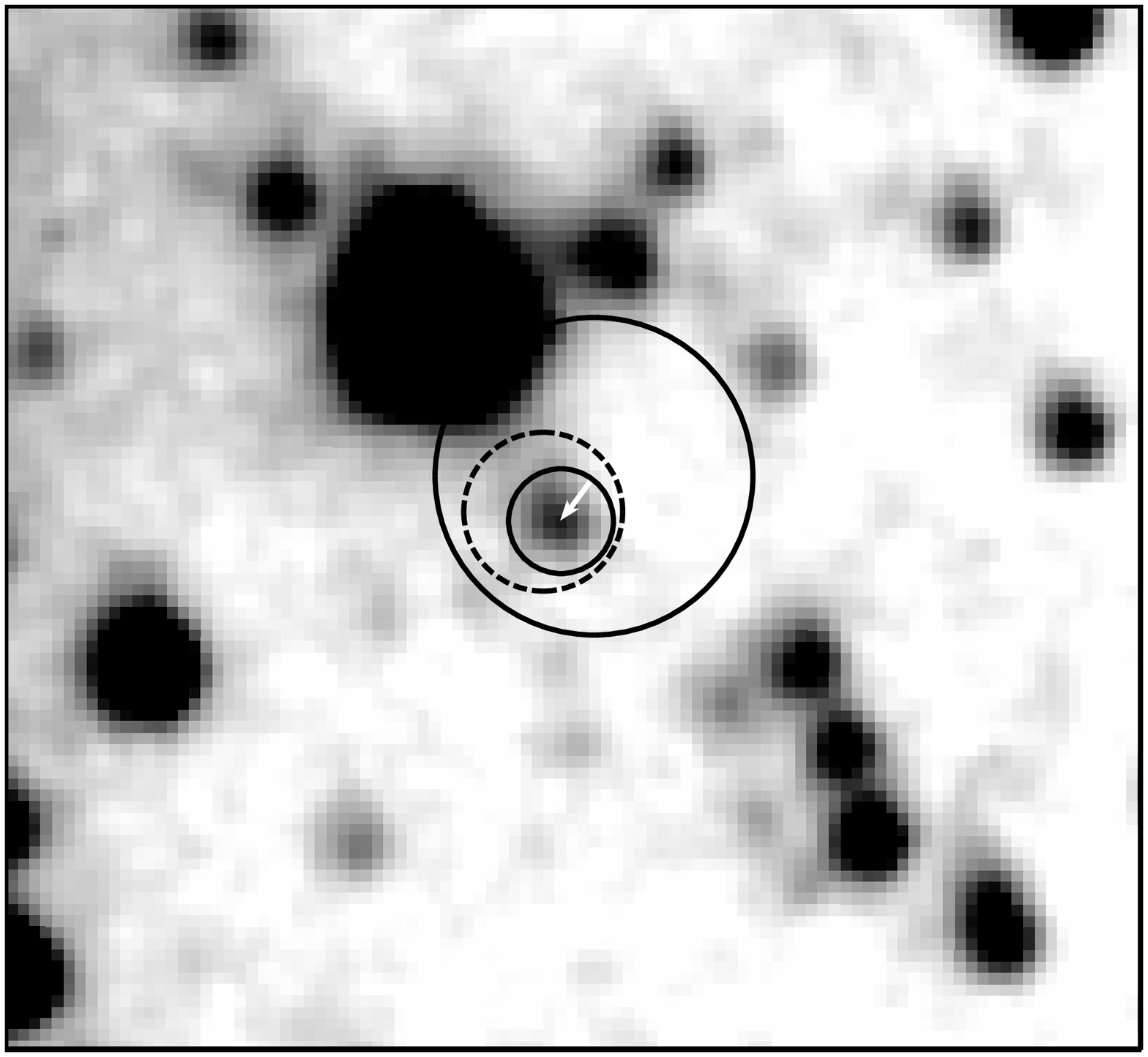}
    \includegraphics[width=0.9\columnwidth]{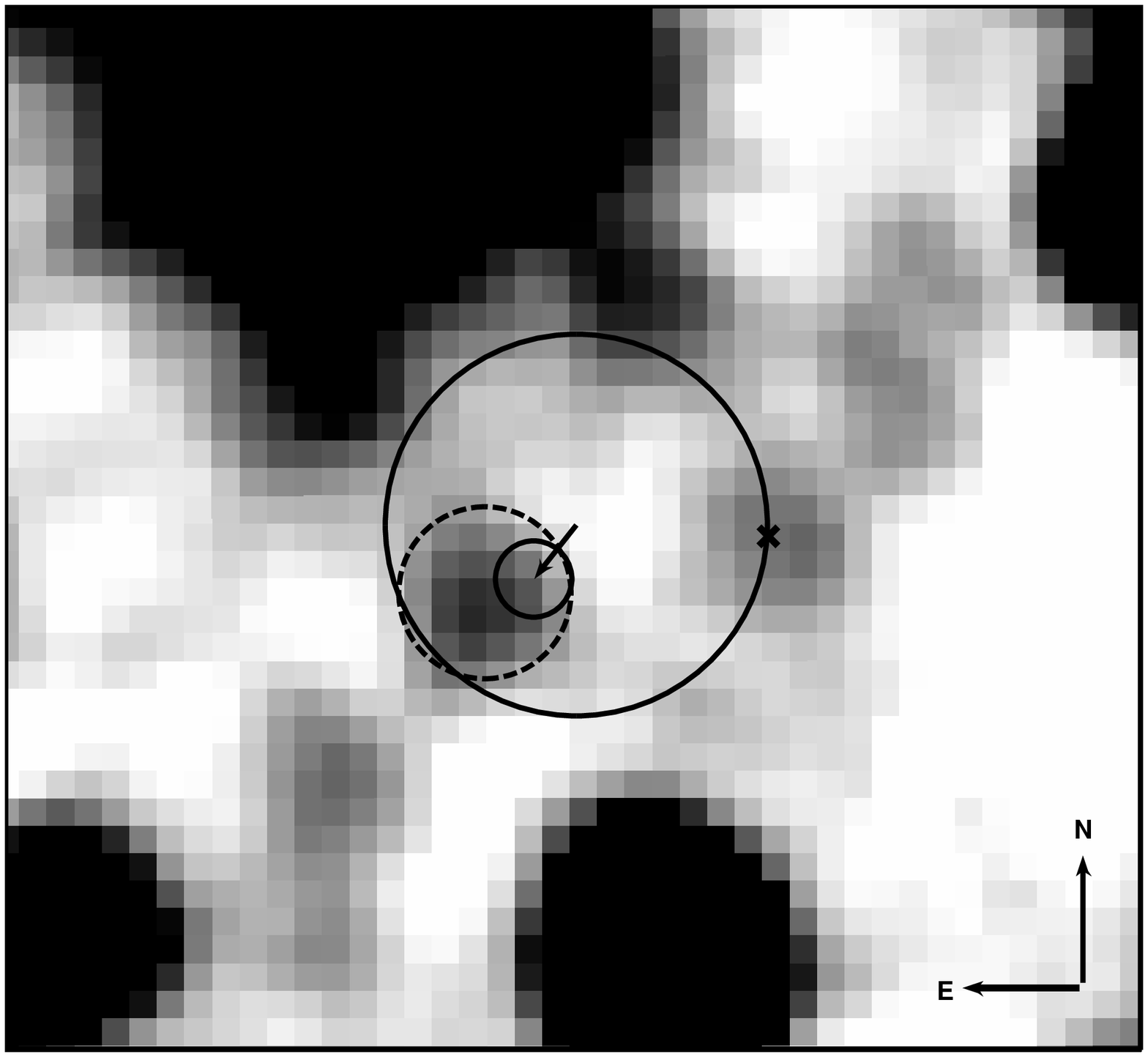}
  }

\caption{SOFI/NTT \emph{Ks}-band images of the field of the sky containing X-ray pulsars AX\,J1749.2$-$2725 (left) and AX\,J1749.1$-$2733 (right). The plate scales are 0.144\arcsec and 0.288\arcsec, respectively. Large solid
circles show the standard (statistical plus systematic) positional
uncertainty from XMM-\emph{Newton}/MOS2 measurements. The proposed
systematic shift of the XMM-\emph{Newton} images relative to IR
ones is shown with arrows. Small solid circles show statistical
uncertainties of XMM-\emph{Newton} measurements only, with the proposed
systematic shift applied. Dashed circles show the Chandra absolute
astrometric uncertainty of the sources position. The position of the star,
proposed by \citet{Zur08} as a counterpart for AX\,J1749.1$-$2733, is marked
by the cross.}  \label{fig:sofi}

\end{figure*}

\section{Identification of IR counterparts for AX\,J1749.2$-$2725 and
  AX\,J1749.1$-$2733}

Coordinates of AX\,J1749.2$-$2725 and AX J1749.1$-$2733 were derived from
XMM-\emph{Newton} data using standard methods described in the SAS manual
(the \emph{eboxdetect} procedure). They have
statistical uncertainties of 0.7\arcsec\ and 0.4\arcsec, respectively. The
additional systematic uncertainty of the XMM-\emph{Newton} pointing
direction can be as large as
$\simeq$2\arcsec\footnote{http://xmm2.esac.esa.int/external/}.

The standard way to minimize this systematic uncertainty is to use other
X-ray sources on the same X-ray image, with clearly identified optical
conterparts. The MOS2/XMM-\emph{Newton} 2--10 keV image of the field of the
sky with pulsars AX\,J1749.1$-$2733 and AX\,J1749.2$-$2725 is shown in
Fig.\ref{maps}. One can see that both sources are observed in the same
XMM-\emph{Newton} pointing and no other X-ray sources are reliably detected
near them. Therefore, in our case the systematic uncertainty of the
XMM-\emph{Newton} astrometry can not be determined directly. But it is
important to note, that this uncertainty should be approximately the same
for both pulsars, because they are both sutuated close to the center of the
detector where the rotation uncertainty is small.

The \emph{Ks}-band SOFI/NTT images of the field of the sky containing each of pulsars are shown
in Fig.~\ref{fig:sofi}. Only one relatively faint star is clearly detected
inside the XMM-\emph{Newton} error circle around AX\,J1749.2$-$2725 with
a small shift relative to the X-ray position (Fig.~\ref{fig:sofi}, left). In
the field of AX\,J1749.1$-$2733 a faint IR star is
also detected approximately in a similar position relative to the center
of the XMM-\emph{Newton} error circle as for AX\,J1749.2$-$2725
(Fig.~\ref{fig:sofi}, right). The shift between the X-ray position of pulsars
and these two stars can be preliminary regarded as a systematic error in the
XMM-\emph{Newton} astrometric solution (in our case it is about 0.7\arcsec).
Therefore, we propose these two stars as IR counterparts to
AX\,J1749.2$-$2725 and AX\,J1749.1$-$2733.

Dashed circles on Fig.~\ref{fig:sofi} show the Chandra uncertainty regions
around pulsars positions, which were obtained from the pipeline data. The
uncertainty in both cases is about $1\arcsec$ and mainly systematic. These
Chandra positions coincide well with two IR sources proposed above based on
XMM-\emph{Newton} data. Thus coordinates of IR counterparts of
AX\,J1749.2$-$2725 and AX\,J1749.1$-$2733 (and consequently pulsars
themselfs) can be determined as: R.A., Dec.:
$17^h\,49^m\,12^s\!.4$,~$-27^\circ\,25\arcmin\,38\arcsec\!\!.34$
(uncertainty $\sim0\arcsec\!\!.1$) and
$17^h\,49^m\,06^s\!.85$,~$-27^\circ\,32\arcmin\, 32\arcsec\!\!.9$
(uncertainty $\sim0\arcsec\!\!.2$), respectively. Note, that the IR
counterpart for AX\,J1749.1$-$2733 is differed from the star suggested by
\cite{Zur08}.It was significantly detected in every of about 400
observations of this field by the SOFI/NTT telescope in the \emph{Ks}-filter
with an approximately persistent magnitude (see Fig.~\ref{fig:sofi}).

\begin{figure}

\begin{center}
\includegraphics[width=0.9\columnwidth]{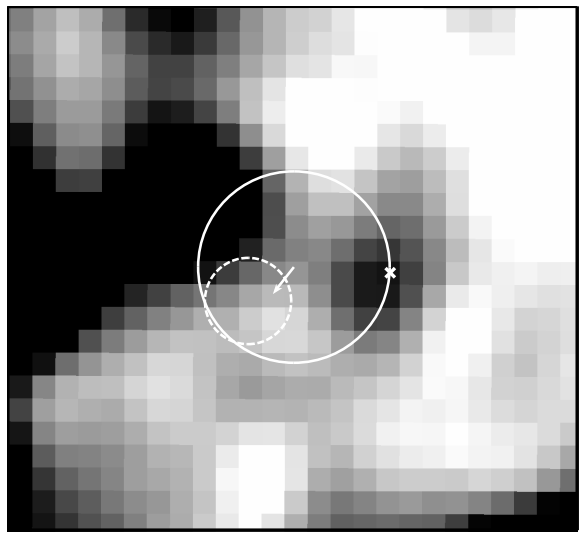}
\end{center}

\caption{The \emph{i\arcmin}-band image of the field of the sky near AX\,J1749.1$-$2733
obtained with the RTT150 telescope. The circle, shift arrow and cross are the same as in
Fig.~\ref{fig:sofi},right).}  \label{fig:i}

\end{figure}

Both proposed counterparts are not detected in optical bands. The one of
AX\,J1749.2$-$2725 is hard to observe due to a very bright nearby star to
the NE from the pulsar (Fig.~\ref{fig:sofi}). The \emph{i\arcmin}-band image of the
AX\,J1749.1$-$2733 field of the sky obtained with the RTT150 telescope is shown in
Fig.~\ref{fig:i}.
Magnitudes and upper limits for
both IR counterparts are summarized in Table~\ref{tab:1}: measurements in
\emph{JHKs} bands were obtained with the SOFI/NTT instrument and the upper
limit in the \emph{i\arcmin}-band was derived from RTT150 data.

\begin{table}\label{tab1}
  \caption{Magnitudes of the proposed counterparts}
  \label{tab:1}
  \centering
  \begin{tabular}{llll}
    \hline
    \hline
    band   & AX\,J1749.2$-$2725 & AX\,J1749.1$-$2733 \\
    \hline
    \emph{i\arcmin}  & --- & $> 20.5$ \\
    \emph{J}  & $18.58\pm 0.21$ & $> 18.7$ \\
    \emph{H}  & $16.57\pm 0.07$ & $17.43\pm0.14$\\
    \emph{Ks} & $14.95\pm 0.05$ & $15.18\pm0.03$ \\
    \hline
  \end{tabular}
\end{table}

\section{Spectral type and distance estimations}

\begin{figure}

\includegraphics[width=\columnwidth,bb=60 200 575
694]{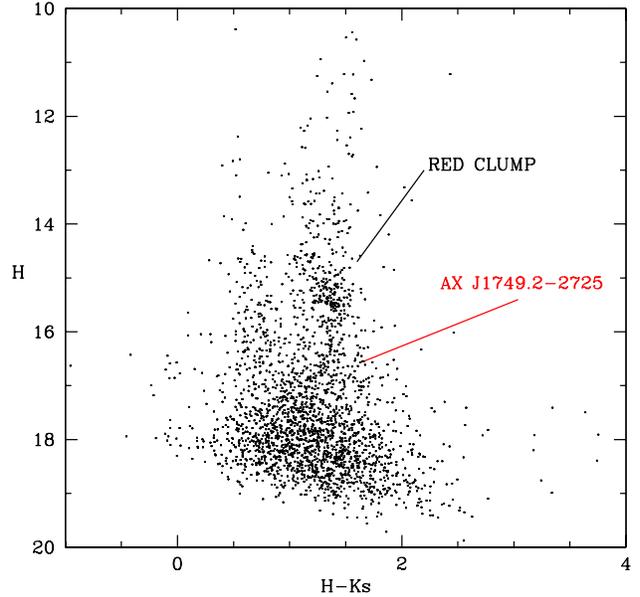}
\caption{Color--magnitude diagram obtained using SOFI/NTT data for the stars
from the $1'\times2'$ sky field near AX\,J1749.2$-$2725.}\label{fig:cm}

\end{figure}

In order to estimate the spectral type of these counterparts we need to
measure the interstellar extinction to the Galactic Center (GC) in the direction
to these objects. Using SOFI/NTT \emph{H} and \emph{Ks} images we constructed the
color--magnitude diagrams (CMD) for all stars in $1'\times2$\arcmin\ box fields
near each of the sources (see Fig.~\ref{fig:cm} for AX\,J1749.2$-$2725). The red
giant branch and, particularly, the red clump giants (RCG) are clearly seen
in this diagram because almost all of them are located
approximately at the same distance, which corresponds to the distance to the
Galactic Center or galactic bulge ($8.4\pm0.4$~kpc, \citealt{Pacz98}).

Using the observed positions of RCG at the color-magnitude diagrams and
their unabsorbed colors and magnitudes, which were found using {\em 2MASS}
IR photometry of RCG from the {\em Hipparcos} catalog, we estimated the
interstellar extinction as $A_{H}=2.1 \pm 0.1$ and $A_{H}=1.42 \pm 0.25$ for
the fields near AX\,J1749.2$-$2725 and AX\,J1749.1$-$2733, respectively. The
extinction law measured from RCG colors turns out to be $A_H/E(H-K) = 1.67
\pm 0.12$ for the both fields and is different from the standard extinction
law in the galactic plane $A_H/E(H-K) = 2.75$ \citep[e.g.,][]{Car89}. The
similar analysis based on the \emph{J-K} color-magnitude diagram also showed
a strong difference between the measured extinction law in the
AX\,J1749.2$-$2725 field $A_J/E(J-K)= 1.29 \pm 0.14$ and standard one
$A_J/E(J-K)= 1.69$. These conclusions are in agreement with results of
\citet{Popow00,Udal03,Rev10}, which demonstrated the significant difference
of the extinction law in the galactic bulge and disk.

Early type luminous stars with the magnitudes similar to that observed for
the AX\,J1749.2$-$2725 counterpart would be located at the distance of the
Galactic center or larger.  It is easy to show if we assume that the studied
system is located at the distance of the Galactic Center and, consequently,
has the extinction of $A_H\sim2.1$. Proposing the different classes of the
stars as a companion of AX\,J1749.2$-$2725 we can obtain their apparent
\emph{H} magnitudes for the above mentioned distance and extinction. If this
magnitude is higher than observed one, that the star of such a type should
be located before GC, in the opposite case -- behind it.  Using the observed
color of the AX\,J1749.2$-$2725 counterpart $(H-Ks)\approx1.6$, and the
extinction law measured from RCG colors above, we can estimate the
extinction for these stars as $A_H\approx2.7$. Late type star with the same
observed magnitude would be located at distance smaller than that of the
Galactic Center and the standard extinction law should be used to estimate
their extinction. In that case from the observed reddening $E(H-K)>1$ we
estimate the extinction $A_H>2.8$.

\begin{figure*}
 {
\includegraphics[width=0.9\columnwidth,bb=60 200 575 694]{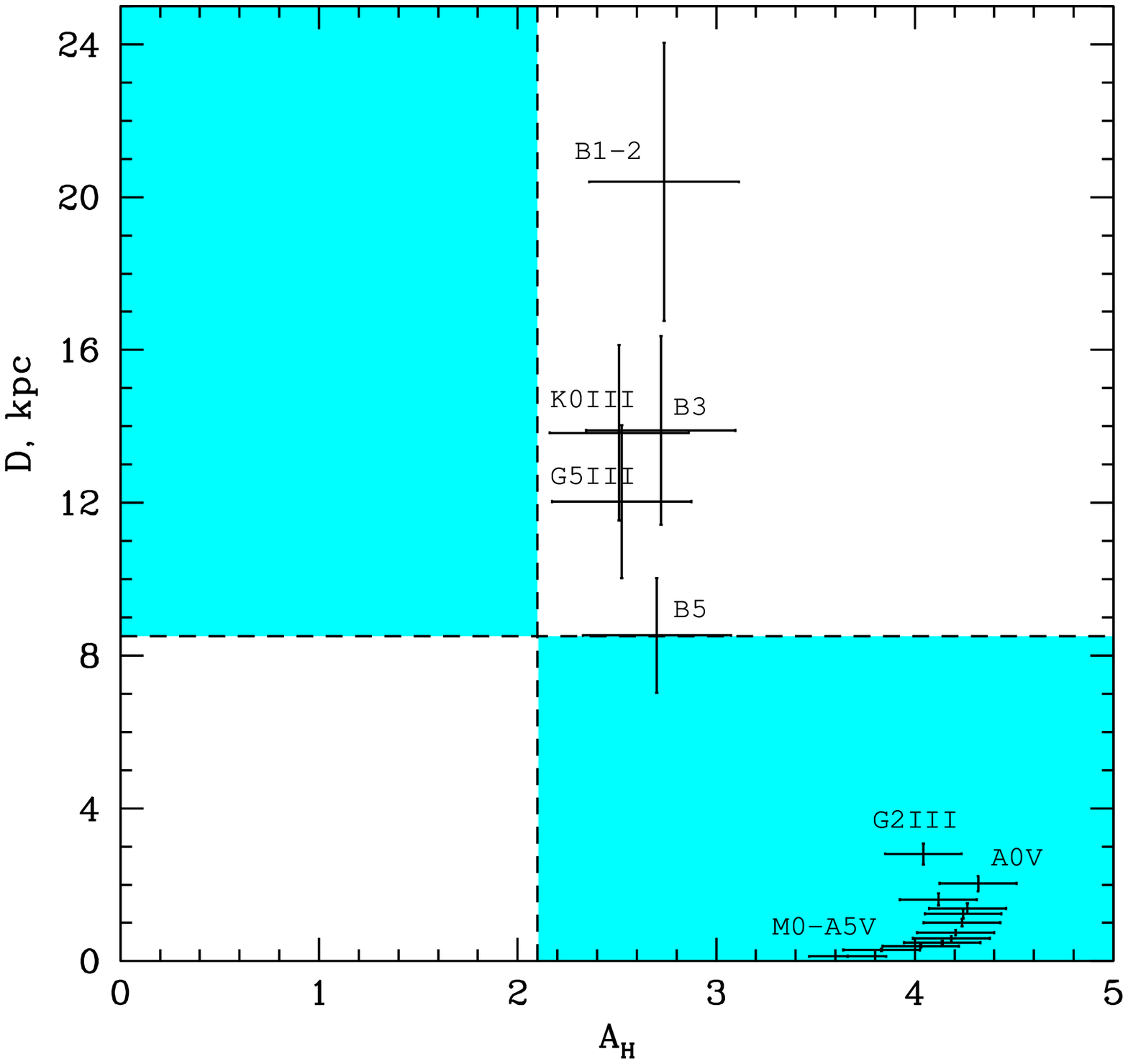}
\includegraphics[width=0.9\columnwidth,bb=60 200 575 694]{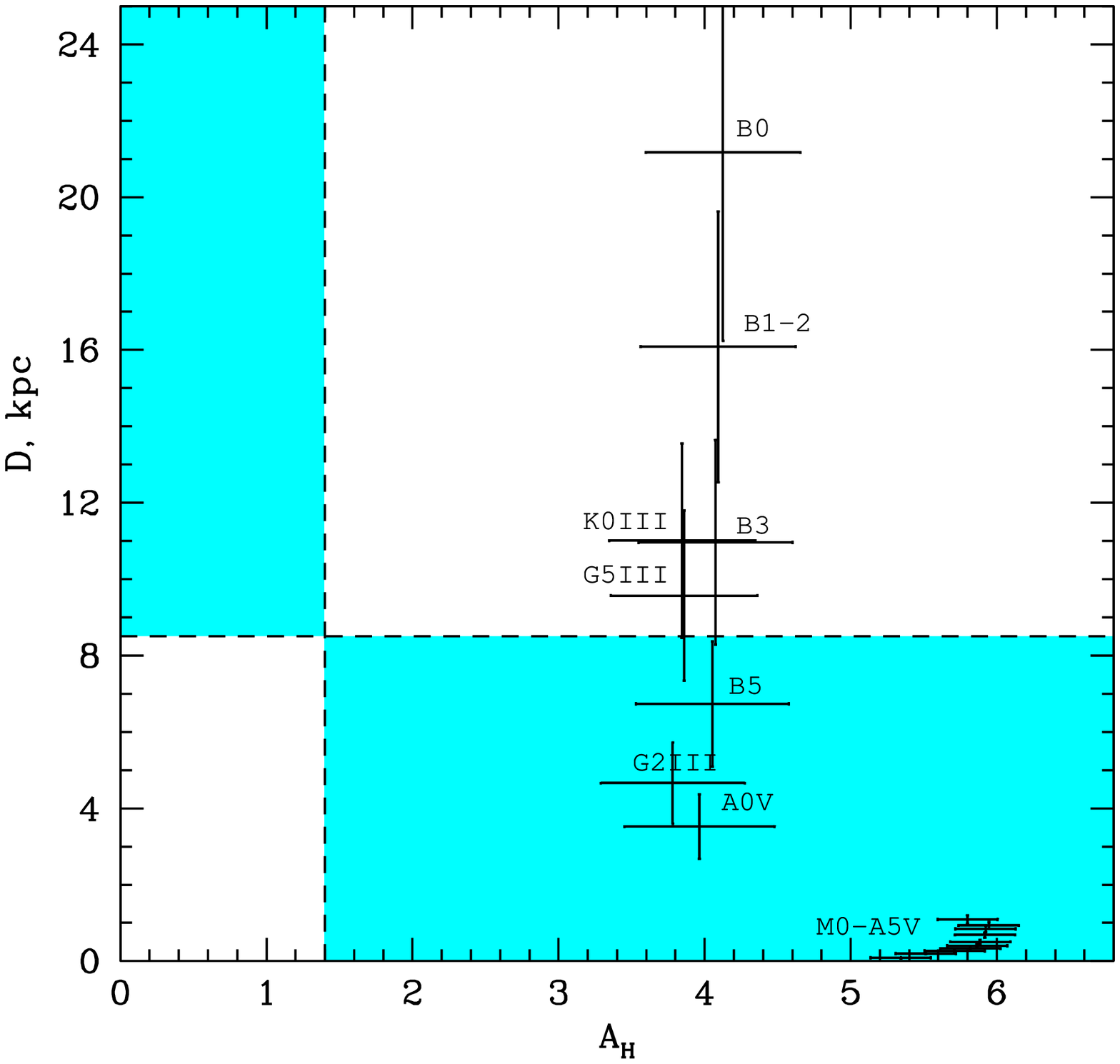}
}

\caption{Extinction -- distance diagram for stars which would be observed
with magnitudes and colors similar to that of the AX\,J1749.2$-$2725 (left)
and AX\,J1749.1$-$2733 (right) IR counterparts. Dashed horizontal lines show
the distance to the Galactic Center; dashed vertical lines indicate the
extinction value to the Galactic Center for both fields. Note, that based on
the available data we could not determine the luminosity class of B$-$stars
unambiguously.}  \label{fig:D_NH}

\end{figure*}

In both cases the extinction is larger than that measured above using RCG in
the galactic bulge. Therefore, AX\,J1749.2$-$2725 should be located at
larger distances, behind the Galactic Center. From the measured magnitudes
of its counterpart we conclude that it should be earlier than B5 if a main
sequence star is suggested. From the other side the main sequence star
with spectral type earlier than B1-2 would be located at the distance larger
than 24 kpc, i.e. outside the Galaxy (Fig.~\ref{fig:D_NH}, left). This
diagram shows the possibility of different classes of stars to be the
counterpart of the studied source in the assumption of an appropriate
absorption law.  Only stars placed on white fields of the diagram may be the
counterpart of the source.  If star is placed on the cyan field it means
that: 1) it would be located before the Galactic Center, but the extinction
would be higher than for the GC; 2) it would be located behind the Galactic
Center, but the extinction would be lower than for the GC. Applying the
similar reasoning for AX\,J1749.1$-$2733 we conclude that it is also located
behind the Galactic Center and the type of its counterpart should be earlier
than B3 and later than B0 for the main sequence stars (Fig.~\ref{fig:D_NH},
right).


Finally note, that giants with spectral types G5-K0 could also fit the
observed magnitudes and colors of the counterpart of the
sources (Fig.~\ref{fig:D_NH}).  This possibility does not looks very
surprising taking into account a recent discovery of a new class of
symbiotic X-ray binary systems, composed of a compact object, most likely a
neutron star, accreting from the wind of an M-type giant \citep[see,
e.g.,][]{Nes10}. However if we assume M0$-$9 giants as the counterpart of
studied sources, that AX\,J1749.2$-$2725 would be located at the distance of
40$-$54 kpc and AX\,J1749.1$-$2733 -- at the distance of 30$-$40 kpc, well
beyond of the Galaxy.

\section{SUMMARY}

Based on the data of XMM-\emph{Newton} and Chandra observatories we
significantly improved localizations and identified infrared counterparts
for two X-ray pulsars AX\,J1749.1$-$2733 and AX\,J1749.2$-$2725, located in
the direction to the Galactic Center. It was shown that the extinction law
towards the galactic bulge is significantly differed from the standard
one. Using this result we estimated the spectral class of the counterparts
for both sources and concluded that they belong to the class of high mass
X-ray binaries located behind the Galactic Center, presumably in one of
spiral arms \citep{lut05}. Most likely candidates to be counterparts seem to
be the B3 star at the distance of $d = 14\pm2.5$ kpc for AX\,J1749.2$-$2725
and the B3 star at the distance of $d = 11\pm3$ kpc or the B1-2 star at the
distance of $d = 16\pm3.5$ kpc for AX\,J1749.1$-$2733. Note, that these
estimations of the distance for AX\,J1749.2$-$2725 are in a good agreement
with the ones obtained from the pulse period changes (see Sect.3).

\section {ACKNOWLEGEMENTS}

Authors thank M.~Revnivtsev for the discussion of methods and results. We
also thank to the anonymous referee for very important and useful comments,
which allow us to improve the paper. This work was supported by the Programs
of the Russian Academy of Sciences ``The origin, structure, and evolution of
objects of the Universe'' and ``Active and stochastic processes in the
Univerce'', grant NSh-5069.2010.2 for support of leading scientific schools
and RFBF grants 07-02-01051, 10-02-01442. We are grateful for the X-ray data
to HEASARC Online Service. A significant part of results was based on
observations made with the European Southern Observatory telescopes obtained
from the ESO/ST-ECF Science Archive Facility.

\label{lastpage}

\end{document}